\documentclass[onecolumn,usenatbib]{mnras}

\usepackage[T1]{fontenc}

\DeclareRobustCommand{\VAN}[3]{#2}
\let\VANthebibliography\thebibliography
\def\thebibliography{\DeclareRobustCommand{\VAN}[3]{##3}\VANthebibliography}

\usepackage{graphicx}	
\usepackage{amsmath}	
\usepackage{amssymb}	

\title[Jitter Radiation for GRB VHE Photons]{Jitter Radiation: towards TeV-Photons of Gamma-ray Bursts}
\author[J. Mao \& J. Wang]{
Jirong Mao$^{1,2,3}$\thanks{E-mail: jirongmao@mail.ynao.ac.cn (JM)},
and Jiancheng Wang$^{1,2,3}$
\\
$^{1}$Yunnan Observatories, Chinese Academy of Sciences, 650011 Kunming, Yunnan Province, China\\
$^{2}$Center for Astronomical Mega-Science, Chinese Academy of Sciences, 20A Datun Road, Chaoyang District, Beijing, 100012, China\\
$^{3}$Key Laboratory for the Structure and Evolution of Celestial Objects, Chinese Academy of Sciences, 650011 Kunming, China
}

\date{Accepted XXX. Received YYY; in original form ZZZ}

\pubyear{2021}

\begin{document}
\label{firstpage}
\pagerange{\pageref{firstpage}--\pageref{lastpage}}
\maketitle

\begin{abstract}
The synchrotron mechanism has the radiation limit of about 160 MeV, and it is not possible to explain the very high energy (VHE) photons that are emitted by high-energy objects. Inverse Compton scattering as a traditional process is applied for the explanation of the VHE emission. In this paper, jitter radiation, the relativistic electron radiation in the random and small-scale magnetic field, is proposed to be a possible mechanism to produce VHE photons. The jitter radiation frequency is associated with the perturbation field. The spectral index of the jitter radiation is dominated by the kinetic turbulence. We utilize the jitter radiation to explain the gamma-ray burst (GRB 190114C and GRB 180720B) VHE emissions that were recently detected by the Imaging Atmospheric Cherenkov Telescopes (IACTs). We suggest that this mechanism can be applied to other kinds of VHE sources.
\end{abstract}




\section{Introduction}
Very High Energy (VHE) astrophysical study is on the astronomical research frontier. Radiation mechanism and particle acceleration are two important aspects in the VHE astrophysical research. Some extragalactic objects, such as gamma-ray bursts (GRBs) and blazars, are high-energy sources, and they are also linked to the cosmic-ray origin.

Imaging Atmospheric Cherenkov Telescopes (IACTs) are usually applied to detect VHE photons, and some great achievements have been obtained on the VHE GRB observation. GRB 190114C is the first GRB detected by the Major Atmospheric Gamma Imaging Cherenkov (MAGIC) telescope in the $0.3-1$ TeV energy band \citep{acciari19}. The multiwavelength observations on this GRB were also performed \citep{magic19b}. The afterglow of GRB 180720B was detected by High Energy Stereoscopic System (HESS; Abdalla et al. 2019). GRB 190829A was also detected by HESS within 4.2 h after the trigger \citep{de19}.

Synchrotron mechanism, which is the radiation as relativistic electrons moving by the gyroradius way in the large-scale magnetic field, is popularly adopted for the explanation of the GRB high-energy emission. However, the radiation has a peak limited to be about 160 MeV, and it is hard to be used for the interpretation of the VHE emission. One can traditionally take inverse Compton (IC) mechanism (synchrotron photons are scattered by external high-energy electrons) or synchrotron self-Compton (SSC) mechanism (synchrotron photons produced by high-energy electrons are scattered by the high-energy electrons themselves) to explain the emission above a few hundred MeV. In particular, the GRB emission detected by the IACTs is usually considered as the result of the IC/SSC mechanism (e.g., Wang et al. 2019; Fraija et al. 2019a; Zhang et al. 2020; however see an early discussion on the synchrotron radiation limit and the photon numbers via IC process by Gao et al. 2009).

We pay attention to some issues related to the IC/SSC mechanism for the explanation of the GRB emission in the high-energy band. First, there are only a few GRBs in which the VHE emissions are detected by IACTs. The GRB properties in the high-energy band can be statistically investigated in the {\it Fermi} observational sample.
In some GeV-detected GRBs, the emissions
in the GeV energy band have harder spectra than the emissions in the MeV energy band. When we consider synchrotron radiation to explain the GRBs in the MeV energy band, the GeV-emissions in these GRBs can be explained by the IC/SSC mechanism \citep{sari01}.
Some GeV-GRBs detected by {\it Fermi}-LAT, such as GRB 080916C, GRB 090510, GRB 090902B, GRB 090926A, GRB 110731A, GRB 130417A, and GRB 160905A, can be well explained by the IC/SSC mechanism \citep{corsi10,he11,peer12,liu13,tam13,fraija15,fraija16a,fraija16b, yassine17}. 
However,
\citet{mao20} comprehensively investigate the GRBs detected by both {\it Fermi}-GBM and {\it Fermi}-LAT. Some GRBs have hard spectra, while some GRBs have soft spectra. The spectral diversities of the {\it Fermi}-detected GRBs cannot simply explained by the IC/SSC mechanism.
Second, for one GRB, the IC/SSC radiation usually has a time delay to the synchrotron radiation.
\citet{ajello19} investigated the GBM emission duration vs the LAT onset time.
In some GRBs, LAT emission occurs a few hundred seconds later to the GBM emission. In the internal shock scenario, the LAT emission produced by IC/SSC mechanism can be delayed, and the delay time can be compared to the flux variability times-cale. The time-scale is about 1 s. However, by the IC/SSC mechanism in the internal shock scenario, it is hard to understand the GRBs with the long delay times of the LAT emissions. Then, one may consider the IC/SSC mechanism in the external shock scenario to explain the LAT emission. Some GRBs have long time duration in the GeV band, and these GRBs have long time delay of the LAT emission onset. In these cases, a very low bulk Lorentz factor is required, and the very low Lorentz factor cannot be used to reproduce the GeV emission \citep{ajello19}. In addition, a suitable number density of the surrounding medium at relatively large fireball radius is also required in the IC/SSC external shock scenario to explain the GRBs having the long duration in the GeV band and the long delay of LAT emission.
Third, synchrotron emission from internal shocks usually has low efficiency. However, the electron energy described by the parameter $\epsilon_e$
and the magnetic energy described by the parameter $\epsilon_B$ are not well constrained \citep{sari98}.
For example, different numbers of $\epsilon_e$ and $\epsilon_B$ provides different hydrodynamic evolution
and radiation properties \citep{mao01a,mao01b}. Furthermore, the radiation processes are complicated
if we consider the magnetically dominated case.
The magnetic energy is expected to have a very quick dissipation. 
The electron cooling is very efficient and the cooling time is very short because of the very large magnetic energy density.
The details on the magnetic energy dissipation are still under debate. 
\citet{gh20} pointed out that the electrons have no completely cooling. It seems that the electron cooling of the GRB emission is a problem for the synchrotron radiation. Consequently, IC/SSC mechanism suffers the same problem. Proton synchrotron was suggested as one possible mechanism of low radiative efficiency for GRB emission. Fourth, the non-thermal process efficiency of the VHE GRB emission was estimated. Synchrotron and IC/SSC mechanisms are very efficient for VHE GRB emission. But proton synchrotron cannot produce enough photons of GRB 190114C in the TeV energy band \citep{acciari19}. In this paper, we consider an alternative radiation mechanism to explain the GRB emission in the TeV energy band.

Recent studies pay attention to the radiation mechanisms that have been linked to the microdynamics. The microdynamics research goes into the issues that focus on the small length-scale properties. The kinetic turbulence has been comprehensively investigated \citep{howes15,ser15}. In particular, different turbulent regimes in fluid and kinetic cases have been well established \citep{sch09}. If particles have strong limits to be accelerated by relativistic shocks, relativistic turbulence is suggested to be efficient for the particle acceleration, and the turbulence occurs at small lengthscales \citep{bell18}. In the magnetized plasmas, particles can be ejected from the reconnecting current sheets and accelerated by the turbulent fluctuation \citep{comisso18}. The synchrotron radiation and the IC/SSC mechanism can be successfully linked to the microdynamics, while we note that the electron cooling of the synchrotron radiation is in the bipolar and large-scale magnetic field.     

Microdynamics is considered at small lengthscales, and high-energy emission usually shows short time-scale duration. We then consider a certain radiation that has a short time-scale in a small length-scale. Jitter radiation, the relativistic electron radiation in the random and small-scale magnetic field, has been proposed \citep{me99,me00,me09}, and it might be associated with the particle acceleration and the magnetic field amplification in some microphysical properties \citep{mizuno14,sironi15,mar16}.
It indicates that the jitter radiation can be applied to high-energy sources. In particular, we have performed a series of analytic work on the jitter radiation. It was the first time that we introduced the turbulent magnetic field to the jitter radiation \citep{mao07}. We attempted the jitter radiation to explain the GRB prompt emission properties \citep{mao11}, and the jitter polarization was also investigated \citep{mao13,mao17}.

We utilize the jitter radiation to explain the GRB VHE emission in this paper, and two characteristics of the jitter radiation are illustrated. First, the spikes in the GRB prompt lightcurve have very short times-cales, and a short time-scale corresponds to a small emission length-scale. The microdynamics mentioned above is extended to the kinetic scale, and the jitter radiation is well constrained in a small length-scale. Thus, the jitter radiation accompanied with the microdynamics in the small lengthscale is suitable to investigate the GRB physics. Second, it seems difficult to use the IC/SSC mechanism to explain the spectral diversities of {\it Fermi}-detected GRBs, but the spectral diversities can be explained by the turbulence cascade in the jitter radiation framework \citep{mao20}. This suggests that the jitter radiation is possible to explain the GRB VHE emission.

It is not necessary to have a tautology to our previous work on the jitter radiation. Here, we focus on the physical issues addressed in \citet{mao07} and \citet{mao11}. A single electron radiation in the small-scale magnetic field can be written as 
$I_\omega=\frac{e^2\omega}{2\pi c^3}\int^{\infty}_{\omega/2\gamma^2_\ast}\frac{\left|\bf{w_\omega'}\right|^2}{\omega'^2}(1-\frac{\omega}{\omega'\gamma^2_{\ast}}+\frac{\omega^2}{2\omega'^2\gamma^4_{\ast}})d\omega'$,
where $\gamma^{-2}_{\ast}=\gamma^{-2}+\omega^2_{pe}/\omega^2$, $\omega'=(\omega/2)(\gamma^{-2}+\theta^2+\omega^2_{pe}/\omega^2)$, $\omega_{pe}=(4\pi e^2n/m_e)^{1/2}$ is the plasma frequency of the background, $\omega$ is the radiation frequency, $\theta$ is the angle between the radiation direction and the electron velocity, $\gamma$ is the electron Lorentz factor, $n$ is the electron number density, $m_e$ is the electron mass, $e$ is the electron charge, and $c$ is the light speed. This is the general presentation of the radiation, and the term $\bf{w}_{\omega'}$ is the Fourier transform of the electron acceleration term in the small-scale magnetic field. One may obtain different radiation results depending on the different $\bf{w_\omega'}$ in detail. We can perform the perturbation theory to the above equation, and
the radiation can be further presented as
$I_\omega=\frac{e^4}{m^2c^3\gamma^2}\int^{\infty}_{1/2\gamma_\ast^2}d(\frac{\omega'}{\omega})(\frac{\omega}{\omega'})^2{(1-\frac{\omega}{\omega'\gamma_\ast^2}
+\frac{\omega^2}{2\omega'^2\gamma_\ast^4})} \int{dq_0d{\bf q} \delta(w'-q_0+{\bf qv})K({\bf q})\delta[q_0-q_0({\bf q})]}$.
We emphasize that
the radiation field is strongly related to the perturbation field. The dispersion relation $q_0=q_0({\bf q})$ is in the perturbation field, and the radiation field is linked to the perturbation field by the relation $\omega'=q_0-{\bf qv}$. Here, we pay attention that $\bf{v}$ is the velocity in the perturbation field.

We present the possibility in Section 2.1 that the jitter photons can reach the TeV energy band. The electron Lorentz factor in the jitter regime is given when we consider the electron radiation.
In Section 2.2, we compare the modeling analysis to the observational results.
In Section 3, we discuss the validation of the kinetic turbulence in the jitter radiation.
A brief conclusion is given in Section 4.

\section{VHE Jitter Radiation and Observational Constraints}
\subsection{VHE radiation of jitter mechanism}
The radiation frequency of the jitter radiation in our scenario is dependent on the relation between the perturbation field and the radiation field.
In principle, the perturbation field can be either fluid field or plasma field. Here, we adopt one example proposed by \citet{mi06} that the structure of the collisionless shock wave transition is in the pair plasmas, and we obtain the dispersion relation $q_0=cq[(1\pm\sqrt{1+4\omega_{pe}/c^2q^2\gamma^2})/2]^{1/2}$, where $\gamma$ is the electron Lorentz factor.
The relativistic electron frequency is $\omega_{pe}=(4\pi e^2n/\Gamma_{sh}m_e)^{1/2}=9.8\times 10^9 \Gamma_{sh}^{-1/2}~\rm{s^{-1}}$, where we take the value $n=3.0\times 10^{10}~\rm{cm^{-3}}$ \citep{me99}, $\Gamma_{sh}$ is the shock Lorentz factor, and we assume $\gamma c^2q^2 \gg \omega_{pe}$. We then consider the dispersion relation to the radiation by the relation $\omega'=q_0-{\bf qv}$. In order to obtain the maximum radiation frequency $\omega'$, we particularly choose the term $\bf {qv}=0$ if the perturbation velocity $\bf{v}$ is perpendicular to the wave number $\bf{q}$. Some complicated cases will be discussed in Section 3.
Thus, we derive the maximum radiation frequency $\omega_{max}=\gamma^2 cq_{max}$.
In our scenario, the turbulent lengths-cale can be derived from the Prandtl number $P_r=10^{-5}T_e^4/n$ by $P_r^{1/2}=q_\eta/q_\nu$, where $T_e$ is the electron temperature,
$q_\eta$ is the maximum length-scale, and $q_\nu$ is the minimum length-scale. When $q_\nu$ is estimated by the length-scale of the turbulent eddy as $q_\nu=2\pi l_{eddy}^{-1}=2\pi (R/\Gamma_{sh}\gamma_t)^{-1}$, where $\gamma_t$ is the turbulent Lorentz
factor \citep{narayan09,mao11}, we obtain the jitter radiation frequency as
\begin{equation}
\begin{aligned}
\omega = 19.6(\frac{n}{3.0\times10^{10}~\rm{cm^{-3}}})^{-1/2}(\frac{T_e}{1.2\times 10^{10}~\rm{K}})^2(\frac{R}{1.0\times 10^{13}~\rm{cm}})^{-1} (\frac{\Gamma_{sh}}{100.0})(\frac{\gamma_t}{10.0})(\frac{\gamma}{1.0\times 10^7})^2~\rm{TeV}.
\end{aligned}
\end{equation}
The electron temperature normalized to the electron mass in the relativistic case can be given by $\Theta=kT_e/m_ec^2$, and $\Theta=2.0$ corresponds to $T_e=1.2\times 10^{10}$ K. It is clearly seen that the TeV-photons of the jitter radiation can be produced by the hot relativistic plasmas, and the GRB prompt photons can be detected by the IACTs when we assume the number density in the internal shock has a number of $3.0\times 10^{10}~\rm{cm^{-3}}$ at a fireball radius of $1.0\times 10^{13}~\rm{cm}$. When we consider the GRB afterglow case (e.g., Sari et al. 1998), we take an example that the GRB surrounding medium has a dense number density of $1.0~\rm{cm^{-3}}$ at a fireball radius of $1.0\times 10^{17}~\rm{cm}$.
We then also obtain the afterglow emission in the TeV energy band as
\begin{equation}
\begin{aligned}
\omega= 34.6(\frac{n}{1.0~\rm{cm^{-3}}})^{-1/2}(\frac{T_e}{1.2\times 10^{10}~\rm{K}})^2(\frac{R}{1.0\times 10^{17}~\rm{cm}})^{-1}(\frac{\Gamma_{sh}}{10.0}) (\frac{\gamma_t}{10.0})(\frac{\gamma}{1.0\times 10^7})^2~\rm{TeV}.
\end{aligned}
\end{equation}
It indicates that the TeV-photons of the GRB afterglow emission can be also produced by the jitter radiation.

The jitter radiation (energy per unit frequency per unit time) produced by a single electron is presented as
\begin{equation}
I_\omega=\frac{4e^4}{3(\zeta_p-1)m_e^2c^{4-\zeta_p}}\gamma^{2(\zeta_p-1)}\omega^{-(\zeta_p-1)}
\end{equation}
when we take $K(q)\sim \int_q^\infty q'^{-\zeta_p}dq'$, and $\zeta_p$ is the turbulent spectral index if the turbulent energy distribution has a power-law shape. Therefore, we obtain the power-law radiation with the index $\zeta_p-1$.

Electron Lorentz factor is a key point to achieve TeV-photons in our scenario. Because each electron has the Lorentz force in the magnetic field, we derive $\omega I_\omega=ecB$ when we adopt the jitter radiation. We then obtain
\begin{equation}
\gamma=[\frac{3(\zeta_p-1)m_e^2c^{5-\zeta_p}B\omega^{\zeta_p-2}}{4e^3}]^{\frac{1}{2(\zeta_p-1)}}.
\end{equation}
The electron Lorenz factor as a function of the magnetic field in the cases of different turbulent spectral indices is shown in Figure 1.
We note that the electron Lorentz factor is constrained by the radiative cooling.

We apply the parameter $\sigma$ that is the ratio between the magnetic energy and the particle energy to identify the magnetization (e.g., Zhang \& Kobayashi 2005) as
\begin{equation}
\sigma=\frac{B^2}{\gamma nm_ec^2}=4.1(\frac{B}{1.0\times 10^6~\rm{G}})^2(\frac{\gamma}{1.0\times 10^7})^{-1}(\frac{n}{3.0\times 10^{10}~\rm{cm^{-3}}})^{-1},
\end{equation}
where the magnetic field and the electron Lorentz factor are constrained by Equation (4). It indicates that the GRB prompt emission is magnetized even a radiative electron has a very high Lorentz factor. If we take a GRB afterglow case as $B=100.0$ G, $\gamma=1.0\times 10^4$ and $n=300.0~\rm{cm^{-3}}$, the magnetization is also satisfied.

\begin{figure}
    \centering
    \includegraphics[width=1.0\textwidth]{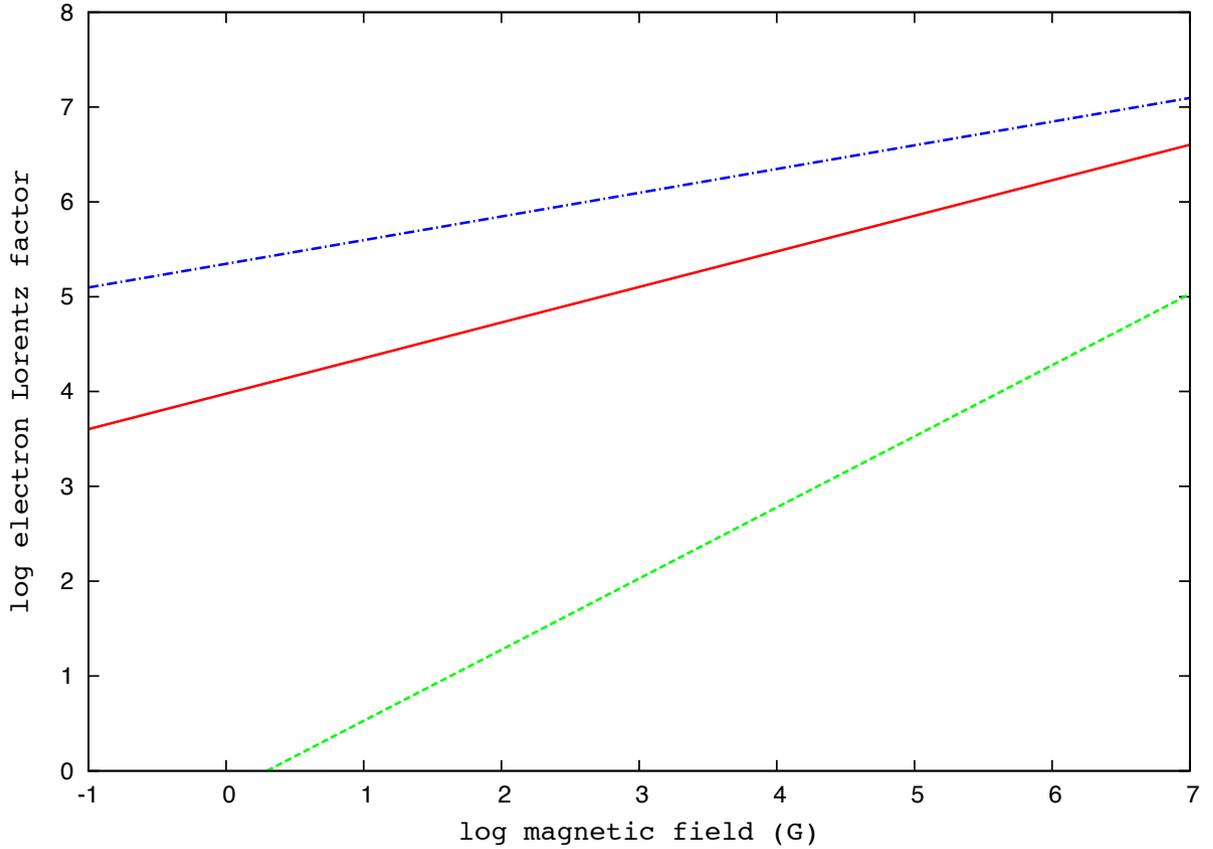}
    \caption{The electron Lorentz factor as a function of the magnetic field when the jitter cooling is considered. The solid line (red) indicates the case with the spectral index $7/3$ of the kinetic turbulence, the dashed line (green) indicates the case with the spectral index $5/3$ of the MHD turbulence, and the dotted line (blue) indicates the case with the spectral index 3 of the collisionless reconnection turbulence.}
    \label{Fig:1}
\end{figure}

We can further estimate the maximum electron Lorentz factor when we do not consider the radiative cooling.
The acceleration distance $L$ can be treated as the GRB fireball thickness $H$. Thus, we estimate $L\sim H\sim R/\Gamma_{sh}^2$, where $R$ is the GRB fireball radius.
We take the electric field $E$ to be the magnetic field. We obtain the electron Lorentz factor and the proton Lorentz factor to be
\begin{equation}
\gamma_{max,e}=eEL/m_ec^2=5.9\times 10^{11}(\frac{R}{1.0\times 10^{13}~\rm{cm}})(\frac{B}{1.0\times 10^6~\rm{G}})(\frac{\Gamma_{sh}}{100.0})^{-2}
\end{equation}
and
\begin{equation}
\gamma_{max,p}=eEL/m_pc^2=3.2\times 10^8(\frac{R}{1.0\times 10^{13}~\rm{cm}})(\frac{B}{1.0\times 10^6~\rm{G}})(\frac{\Gamma_{sh}}{100.0})^{-2},
\end{equation}
respectively. We note that electrons and protons are collisionless.

If we consider the electron radiative cooling, we have the electron Lorentz factor $\gamma$ that should be smaller than the maximum Lorentz factor $\gamma_{max,e}$.
For example, from Equation (4), if we
take the spectral index of the turbulence to be 7/3 and the magnetic field strength to be $1.0\times 10^6$ G,
we obtain the electron Lorentz factor $\gamma=1.0\times 10^6$ in the radiation energy of 1 TeV. If we do not consider radiative cooling, electrons can be accelerated even beyond PeV energy range as presented in Equation (6). This indicates that the electron acceleration is effectively quenched by the radiative cooling. On the other hand, because $\gamma_{max,e}=eEL/m_ec^2=\beta_EL/r_L$, where $\beta_E=E/B=1$, it is possible that the electrons with large Lorentz factors can be still beyond the Larmor radius and escape even we consider the radiative cooling (Here, we simply present $\beta_E=E/B$, such that we make $E$ and $B$ have same units.). Protons can be accelerated to $\gamma\sim 10^8$ if we do not consider the proton radiation. Therefore, we provide an acceleration process that can be applied to the cosmic-ray origin. The above results are derived under two assumptions. First, it is in the magnetically dominated case, and we take Equation (5) to examine this case. Second, we simply assume $\beta_E=E/B=1$, and this yields $E=B$.
Furthermore, 
the electron cooling length-scale is $l_{cool}=\gamma mc^2/eB$ when we consider $E=B$. This is the Larmor radius of a relativistic electron. 
We emphasize that $\gamma$ is related to the cooling length-scale $l_{cool}$, while $\gamma_{max}$ is related to the system lengthscale L. We note that the jitter radiation in the sub-Larmor scale might be also valid \citep{me11}.

\subsection{Jitter radiation compared to VHE GRB observations}
The jitter radiation of a single electron presented in Equation (3) provides a power-law spectral slope, and the slope can be determined by the kinetic turbulence when we consider the kinetic length-scales. The gross jitter radiation that is the result of collecting the radiation contribution from all electrons has the same spectral slope.
From the theoretical point of view, the spectral index $\zeta_p=7/3$ of the turbulence can be obtained from some mechanisms at small length-scales. For example, the Alfv\'{e}nic turbulence at the hydrodynamic scale can go into the kinetic scale to be the kinetic Alfv\'{e}n turbulence, and the number $7/3$ is the typical spectral index of the kinetic turbulence \citep{sch09,zhao16}. From the observational point of view, GRB 190114C detected by the MAGIC telescope has a power-law photon index $2.22^{+0.23}_{-0.25}$ \citep{acciari19}. GRB 190014C was also detected by {\it Fermi}-GBM with a photon index $2.10\pm 0.05$ and by {\it Fermi}-LAT with a photon index $2.02\pm 0.95$, and a photon index of this GRB provided by {\it Swift}-BAT is $2.19^{+0.39}_{-0.19}$ \citep{magic19b}. The photon indices, despite of the relatively large error bars, are very close to the number $7/3$ that is determined by the kinetic turbulence. We suggest that the VHE emission of GRB 190114C might be originated from the jitter radiation with the kinetic Alfv\'{e}n turbulence.

GRB 180720B detected by HESS has a power-law photon index $1.6\pm 1.2$ (statistical) $\pm 0.4$ (systematic), and the {\it Fermi}-LAT observation provided a power-law photon index $2.10\pm 0.10$ to GRB 180720B \citep{abdalla19}. It seems that the spectral index $7/3$ of the kinetic turbulence is hard to be applied to explain the observed photon index given by HESS because of the large statistical error from the observation. The GRB 180720B emission detected by H.E.S.S is considered as the GRB afterglow, and we believe that the jitter radiation in our scenario is also suitable to explain GRB afterglow cases in general (see also an early discussion on the jitter radiation to treat the GRB afterglow case by Workman et al. 2008). We expect that future IACT detection can collect enough VHE photons to precisely obtabin GRB spectrum.

The fluxes of GRB 190114C in $0.3-1$ TeV are $5\times 10^{-8}~\rm{erg~cm^{-2}~s^{-1}}$ and $6\times 10^{-10}~\rm{erg~cm^{-2}~s^{-1}}$ at 80 s and $10^3$ s after the trigger, respectively. The maximum isotropic luminosity can reach about $10^{49}\rm{erg~s^{-1}}$ \citep{acciari19}. In this paper, we can calculate the VHE flux produced by the jitter radiation and compare to the MAGIC observational results of GRB 190114C. We perform integral calculation to Equation (3) in the energy range between $0.3-1$ TeV. The electron energy distribution is assumed to be a power-law, and we put the power-law index to be 2.2. The radiation volume can be estimated as $4\pi R^2H$, where R is the fireball radius, $H\sim R/\Gamma^2$ is the fireball thickness, and $\Gamma$ is the bulk Lorentz factor. The radiation in the TeV energy band can be achieved if we have enough high-energy electrons.
For example, when we assume that GRB VHE emission is originated from the prompt emission, we obtain the jitter radiation flux of $4.9\times 10^{-9}~\rm{erg~cm^{-2}~s^{-1}}$ in the case that the electron number density is $n=3.0\times 10^{11}~\rm{cm^{-3}}$ and the maximum Lorentz factor is $\gamma_{max}=1.0\times 10^6$. The fireball radius is $R=1.0\times 10^{13}$ cm, and the bulk Lorentz factor is $\Gamma=100.0$.
We can also consider that the emission in the TeV energy band is originated from the GRB afterglow.
We obtain the flux of $1.6\times 10^{-8}~\rm{erg~cm^{-2}~s^{-1}}$ in the case that the electron number density is $n=3.0~\rm{cm^{-3}}$ and the maximum Lorentz factor is $\gamma_{max}=1.0\times 10^6$. The fireball radius is at $R=1.0\times 10^{17}$ cm, and the bulk Lorentz factor is $\Gamma=50.0$.
We can further estimate the GRB magnetization by Equation (5). In the two examples mentioned above, the processes are magnetized.

The flux of GRB 180720B observed by HESS is about $5.0\times 10^{-11}~\rm{erg~cm^{-2}~s^{-1}}$ \citep{abdalla19}. We can calculate the VHE flux of the jitter radiation for GRB 180720B.
We obtain the flux of $5.8\times 10^{-11}~\rm{erg~cm^{-2}~s^{-1}}$ in the case that the electron number density is $n=0.5~\rm{cm^{-3}}$ and the maximum Lorentz factor is $\gamma_{max}=1.0\times 10^6$. The fireball radius is at $R=1.0\times 10^{17}$ cm, and the bulk Lorentz factor is $\Gamma=20.0$. It is clear that the process is magnetized in this case.

It is assumed that GRB 180720B has the VHE afterglow. Although multiwavelength observations can provide more information on the GRB bulk dynamics and the prompt emission/afterglow properties \citep{an19,deugarte19,misra19,jor19,ravasio19}, we focus on the GRB VHE emission in this paper. The systematical analysis on the GRB multiwavelength observation is expected to further constrain theoretical models in the future.

\section{Discussion}

The condition $a=eB/q_\eta mc^2<1$ should be satisfied for the jitter radiation, and $a$ is called as the wiggler number \citep{sironi09}. It indicates the deflection angle of the electrons in the random magnetic field compared to the beaming angle. If the deflection angle is smaller than the beaming angle, we have the condition $a<1$, and jitter radiation is valid \citep{me00}. We then calculate the wiggler number as
\begin{equation}
a=\frac{eB}{q_\eta mc^2}=1.0(\frac{B}{1.0\times 10^6~\rm{G}})(\frac{q_\eta}{5.9\times 10^2~\rm{cm^{-1}}})^{-1}.
\end{equation}
The wiggler number is less than 1 if $q_\eta>5.9\times 10^2~\rm{cm^{-1}}$. Here, we take the numbers in Section 2.1 and obtain $q_\eta=1.6\times 10^3~\rm{cm^{-1}}$. Thus, the jitter radiation is valid. In order to achieve the jitter radiation in the TeV energy band, we set $\Theta=kT_e/m_ec^2\ge 2.0$ in this paper. It is normal in GRB cases that the plasmas have the relativistic temperature. Furthermore, \citet{me11} even extended jitter radiation in the large reflection angle case to be $1<a<\gamma$. The jitter radiation still has a power-law spectrum in this regime. Therefore, we believe that the radiation process presented in this paper is fully valid for the investigation of the GRB VHE emission.

We propose the possibility that the jitter radiation can produce the photons in the TeV energy band. In our scenario, we use the shock wave transition properties in the relativistic collisionless shock framework provided by \citet{mi06}, and we assume ${\bf qv}=0$ in the relation between the radiation field and the perturbation field. However, the physical processes related to the perturbation field that take effects on both the radiation frequency and the radiative spectral index are complex. In principle, some important plasma properties should be included. For example, it is well known that the MHD shear Alfv\'{e}n wave has the condition to hold ${\bf qv}=0$. The kinetic Alfv\'{e}n wave (KAW) dispersion relation was comprehensively studied by \citet{ly96}. The kinetic turbulence with the typical spectral index $7/3$ was applied \citep{sch09,zhao16}, and we note that a steeper index $8/3$ was also introduced \citep{bold12}. Although the kinetic Alfv\'{e}n turbulence is taken as one example in our scenario, some other magnetic plasma effects in the relativistic case can be explored. For example, the relativistic kinetic turbulence was numerically simulated, and a very steep turbulent spectral index at sub-Larmor radius was given \citep{comisso18}. We will collect observational data in the different energy bands and investigate different kinetic processes at different lengthscales in the jitter radiation framework as a consequent work in the future.

In order to identify the GRB magnetization, we usually compare
the magnetic field energy to the relativistic electron energy as presented in Equation (5). It is obvious that the electrons with low number density and small Lorentz factor have large magnetization number when we fix the magnetic field strength. However, the number density is related to the Lorentz factor because electrons have an energy distribution that is usually presented by a power-law shape. In Equation (5), we assume that the electrons with the Lorentz factor of $1.0\times 10^7$ have the number density of $3.0\times 10^{10}~\rm{cm^{-3}}$. This case can be the lower-limit of the magnetization. In fact, some electrons with lower Lorentz factor or lower number density have larger magnetization numbers. This indicates that GRB prompt emission is magnetically
dominated if we take $B=1.0\times 10^6$ G. When we consider the VHE emission originated from the GRB afterglow, the cases are also magnetically dominated
if the magnetic field strength is larger than 100.0 G and the number density is smaller than 300.0 $\rm{cm^{-3}}$. 
This result is for all the electrons with $\gamma >1.0$. 
It seems that most cases for both prompt and afterglow emissions can be magnetically dominated.

It is shown from Equation (4) that the electron Lorentz factor is related to the magnetic field. From Equations (6) and (7), we see that the particle acceleration is determined by the electromagnetic field. Therefore, the high-energy electron production prefers a strong magnetic field in this scenario. Here, we further note this issue. The IC/SSC mechanism to reproduce VHE emission of GRB 190114C is in the case of $\epsilon_e\gg \epsilon_B$. It means that the physical process is not dominated by the magnetic field \citep{magic19b}.
The shock acceleration as a general particle acceleration process can produce high-energy electrons.
However, in our scenario, the process is magnetized, and we do not consider the shock acceleration to generate relativistic electrons.
If the magnetic-dominated case is inclined,
some physical mechanisms, such as magnetic reconnection and turbulence in the kinetic scales, can be involved. Therefore, the jitter radiation accompanied with the dynamical processes like kinetic turbulence and kinetic magnetic reconnection occurred in the small lengthscales, can be generally realized for the GRB energy dissipation.

We present some recent tasks on the physical processes dominated by the magnetic field.
Magnetic reconnection can be calculated as a traditional megnetohydrodynamics (MHD) process. Turbulent reconnection was firstly mentioned by \citet{la99}, and it was applied to the relativistic plasmas \citep{taka15}. Stochastic acceleration is involved in the turbulent reconnection \citep{piso18}. It is possible to use the turbulent reconnection to explain GRB phenomena \citep{la19}. 
Furthermore, the spectral feature of the kinetic turbulence in the collisionless reconnection was recently analyzed, and the spectral index has a range from $8/3$ to 3 \citep{lo17}.
In particular, 
\citet{ze16} comprehensively investigated the acceleration site and the particle motion nearby the reconnection region. Particles can be effectively accelerated by the electromagnetic field during the collisionless reconnection process.
Although the complete process of the particle acceleration in the reconnection region is complicated, it could be very interesting that the turbulent properties in the collisionless reconnection can be further investigated with the combination of the jitter radiation.

The dynamical time-scale of the turbulent eddy is estimated by \citet{mao20}.
The eddy turnover time can be estimated by $t_{\rm{eddy}}= (\lambda L)^{1/2}/V_A$, where $V_A\sim c$ is the
Alfv\'{e}n
speed.
Here $\lambda=R/\Gamma\gamma_t$ is the turbulent eddy lengthscale, $L=R/\Gamma^2$ is the outer scale of the turbulence, $R$ is the GRB fireball radius, $\Gamma$ is the bulk Lorentz factor, and $\gamma_t$ is the Lorentz factor of the turbulence. The observed time-scale is the intrinsic time-scale divided by the bulk Lorentz factor $\Gamma$. We assume a thin shell expanding, and the bulk Lorentz factor decreases as $\Gamma=\Gamma_0(R/R_0)^{-\alpha}$ \citep{sari97}.
Thus, we obtain
\begin{equation}
t_{\rm{eddy}}=1.1\times 10^{-3}(\frac{R}{1.0\times 10^{13}~{\rm{cm}}})^{(5\alpha+2)/2}~\rm{s}.
\end{equation}
Here, we take $R_0=1.0\times 10^{13}$ cm and $\Gamma_0=100.0$. The turbulence turn over time is extremely increased, if the turbulence develops as the fireball shell expands.

An expanding shell to produce GRB temporal structure provides a time-scale of $R/\Gamma^2c$ \citep{fen96}. We obtain the
timescale to be $3.3\times 10^{-2}$ s when we take $R=1.0\times 10^{13}$ cm and $\Gamma=100.0$. It is usually assumed that GRB prompt emission is produced by multiple expanding shells, as the spikes shown in prompt emission lightcurves are less than $10^{-3}$ s \citep{bhatt12,golk14}. This indicates that the electron cooling and acceleration time-scales are also very short. However, the duration of the VHE emission in GRB 190114C is about $10^3$ s. It is suggested that the electrons to produce VHE emission can be accelerated many times during the turbulent cascade process \citep{sobacchi20}. From the calculation by Equation (9), the eddy turn over time of the turbulence is $1.0\times 10^3$ s, which is in agreement with the observing duration of the VHE emission in GRB 190114C. This result corresponds to the turbulence developing from $R=1.0\times 10^{13}$ cm to $R=1.8\times 10^{14}$ cm. Here, we take $\alpha=3/2$ and $\gamma_t=10.0$.

It is well known that the synchrotron radiation produced by a single electron with a fixed Lorentz factor has a relatively narrow
frequency range, and the peak frequency is $0.45\gamma^2\nu_L$, where $\nu_L$ is the Larmor frequency. When we sum up the contributions from all the electrons by an electron energy distribution, we can obtain the gross synchrotron radiation that has a relatively large frequency range.
However, in our scenario, the jitter radiation produced by a single electron with a
fixed Lorentz factor has a frequency of $\omega= \gamma^2cq$, and it is determined by the turbulent cascade. Because the turbulent cascade covers a large range in dynamical scale, the jitter radiation of a single electron can have a large frequency range.
When we sum up the contributions from all the electrons by an electron energy distribution, the gross jitter radiation has a single component that extends over vast frequency scales. We further note that sometimes a GRB spectrum shows different photon indices in a wide energy range.
The radiation including two components is usually considered.
In this paper, we note that the jitter radiation as a single component can be used to explain different spectral indices in one GRB. In our scenario, the spectral index of the jitter radiation is determined by the spectral index of the turbulence. When the turbulent cascade develops, GRB spectrum turns out to be softer towards a higher energy band. When the inverse turbulent cascade develops, GRB spectrum turns out to be harder towards a higher energy band. The GRB spectral diversity and the turbulence cascade have been illustrated in detail \citep{mao20}. Thus, we think that different spectral indices shown in a wide energy range can be explained by a single radiation component.

Some GRBs with hard spectra in the GeV energy band can be explained by the IC/SSC mechanism, while it is
hard to explain some GRBs with soft spectra by the IC/SSC mechanism. It is possible to solve the GRB spectral
diversity by the jitter radiation with the turbulent cascade scenario \citep{mao20}. The IC/SSC mechanism
can produce the emission that has a time delay of less than 1 s to the emission produced by synchrotron
radiation, but it is hard to explain the time delay longer than 1 s. Because turbulent cascade requires a
time interval to reach a full development, we suggest that the jitter radiation with the turbulent cascade
can produce the emission with the time delay less than 1 s. It is also possible to explain the emission
with the time delay larger than 1 s by the jitter radiation
with the turbulent cascade, if we consider the emission is at a large fireball radius and GRB has a
relatively small bulk Lorentz factor. The details were presented in \citet{mao20}. In the magnetic-dominated
case, particles are not effectively accelerated by relativistic shocks. Furthermore, as presented in
Section 1, it seems that the electron cooling by both synchrotron and IC/SSC mechanisms are not sufficient
to dissipate magnetic field energy in a short time-scale. \citet{gh20}
suggested a continuous reacceleration for electrons to halt the fast cooling rate.
We note that a part of the magnetic energy released by the magnetic reconnection can be transferred to the electron kinetic energy.
It is important to note that the turbulence cascade in the small lengthscales can be induced by the magnetic reconnection \citep{franci17}. The particle acceleration in the kinetic length-scale was further investigated in the magnetically dominated plasmas when the turbulence is well developed \citep{comisso18}. These works encourage us to further consider the radiation effects. In our opinion, magnetic field energy can
be dissipated by both magnetic turbulence and magnetic reconnection heating at small
length-scales. Particles can be accelerated by turbulence and magnetic reconnection. In the meanwhile, electrons have effective cooling by jitter radiation. The systematical analysis could be performed in detail in the future.

We proposed jitter radiation to explain GRB prompt emission \citep{mao11}.
In that paper, we clearly predicted the maximum
radiation frequency of $10^{11}$ to $10^{18}$ eV if fireball radius is from $10^{13}$ to $10^{16}$ cm.
In such cases, we took the magnetic field to be $10^6$ G and the bulk Lorentz factor to be 100 
(see the details presented in the second and third paragraphs of Section 3.3 in Mao \& Wang 2011).
Thus, the prediction of the jitter radiation to the TeV GRBs was preliminarily given. In this paper, we comprehensively apply the jitter radiation to explain the GRBs detected in the TeV energy band.

It is still relatively difficult to distinguish jitter and synchrotron spectra by observations.
If electron energy distribution is assumed to be a power law, both jitter and synchrotron produce
a spectrum with a power-law shape. The synchrotron spectral index is determined by the spectral index
of the electron energy distribution, while the jitter spectral index is determined by
the spectral index of the turbulence. But the observed GRB spectral index has large diversities \citep{mao20}.
We hope that the observations from both Fermi and MAGIC/HESS can provide more samples to constrain
theoretical models\footnote{For example, GRB 201015A was detected by MAGIC recently \citep{blanch20}.}.

Multiwavelength observation provides a systematical study on the GRB radiation mechanism. In particular, {\it Fermi}-LAT
takes an important role on the GRB observations in the GeV energy band. Here, it is helpful to discuss the GRB emission in the GeV energy band for
GRB 1901114C and GRB 180720B.   
GRB 190114C detected by {\it Fermi}-LAT has a short-time flaring feature within 10 s after the trigger.
\citet{fraija19b} suggested that the feature can be explained by the SSC mechanism in the reverse shock regime.
The general GRB radiation process in the reverse shock regime has been investigated \citep{wang01,veres12}. It is found that this short-time brightening feature has been also shown in some other GRBs in the {\it Fermi}-LAT catalog \citep{ajello19}. 
GRB 090510 and GRB 130427A are two examples \citep{fraija16a,fraija16b}. Recently, \citet{fraija20} comprehensively studied
the GRB flaring feature detected by {\it Fermi}-LAT, and the SSC mechanism in the reverse shock regime
can be successfully applied to explain the feature.  
In this paper, we utilize the jitter radiation process in the kinetic turbulence framework to explain the GRB emission in the
TeV energy band. 
The jitter photons can be scattered by the relativistic electrons that produce the jitter photons themselves. The process is very similar to the SSC mechanism, and it is called jitter self-Compton (JSC) mechanism \citep{mao12}. The SSC mechanism uses the photons produced by the synchrotron radiation to be the seed photons, and the JSC mechanism uses the photons produced by the jitter radiation to be the seed photons. If both the jitter radiation and the synchrotron radiation can successfully produce the same seed photons, the JSC and SSC mechanisms can produce the same photons of the bright peaks in the GeV energy band. 
Here, we further note that the successful JSC mechanism is dependent on the maximum electron Lorentz factor.  The maximum number that the relativistic electrons can be accelerated is dependent on the particle acceleration. We propose that the kinetic turbulence takes an important effect on the particle acceleration. In this paper, the power-law index of the electron energy distribution that we adopted to calculate the fluxes of GRB 190114C and GRB 180720B is originated from the results of the kinetic turbulence acceleration.
Some electrons should reach the Lorentz factor of $1.0\times 10^6$ \citep{mao12}. We realize in Equation (6)  that the relativistic electrons can be effectively accelerated.
Thus, the JSC mechanism can be applied to explain the brightening feature in the GeV energy band for some GRBs.

The {\it Fermi}-LAT data of GRB 180720B were obtained in the time interval of $10-630$ s after the trigger.   Some special characteristics in the GeV energy band were identified \citep{fraija19c}. The first high energy photon at 100 MeV was detected about 20 s after the trigger. 
The highest energy photon at about 5 GeV was shown about 150 s after the trigger. The photon density has an extremely increasing feature for a time longer than 50 s. 
We agree that the high-energy emission obtained by {\it Fermi}-LAT is GRB afterglow.  In the fireball model,
SSC mechanism takes a certain time to have the onset of the afterglow. 
In this paper, we suggest that the turbulent turn over time can be applied to estimate both the duration of the prompt emission and the onset of the GRB high-energy afterglow. The increasing photon density observed by {\it Fermi}-LAT indicates the process of the turbulent development in the GRB afterglow stage. From Equation (9), we estimate that the afterglow at the fireball radius of $1.0\times 10^{14}$ can have the turbulent turn over time of about 62 s. This turn over time-scale could be used to estimate the increasing photon density of GRB 180720B in the GeV energy band. In addition, because the turbulent turn over time is dramatically increased when the fireball radius is increased, we speculate that the photon density increasing time is tightly related to the turbulent turn over time-scale.  
When the turbulence has a full development, the photon density increasing can be stopped. The turbulent cascade may further go on after the turbulence is fully developed.  We note that the LAT lightcurve can be fitted by a broken power-law. The two indices are $1.49\pm 0.12$ and $3.09\pm 0.64$, respectively \citep{fraija19c}. If the turbulence turns to be weaker during the turbulent cascade, the jitter/JSC radiation may turn to be weaker as well. Thus, the steepening feature shown in the light-curve may be induced by the turbulence cascade changing.  

Multiwavelength observations by both IACTs and {\it Fermi} satellite may provide some special cases. For example, GRB 190829A was clearly detected by HESS but not detected by {\it Fermi}-LAT \citep{chand20}. \citet{fraija21} and \citet{zhang21} suggested that SSC or IC mechanism produces a detectable flux density for HESS in 80 GeV but a very low flux density in 100 MeV that cannot be observed by {\it Fermi}-LAT. Thus, this peculiar phenomenon of GRB 190829A can be explained. In our model, the jitter radiation with the turbulent feature can reproduce the TeV emissions of  GRB 190114C and GRB 180720B.
Here, we note that the turbulent energy dissipation has the intermittent feature both in space and in time \citep{alex18}. If the energy spectrum of the turbulence has the intermittent feature, the corresponding intermittent feature in the jitter radiation spectrum can be also shown. The intermittent feature in the turbulent flow of GRB 190829A may induce the intermittent feature of the high-energy emissions in the different energy bands. Thus, the detections by {\it Fermi}-GBM and HESS and the non-detection by {\it Fermi}-LAT can be explained by the turbulent intermittency. The intermittent current sheets in the kinetic turbulence were investigated, and it may have an important application in high-energy astrophysical objects \citep{zhd20}. However, the intermittent property is not universal in turbulent flows, and it is strongly dependent on the
space dimension, coherent structure and statistical fluctuations in a certain turbulent flow. How to link turbulent intermittency to radiation phenomena in high-energy objects is a challenge. We shall further explore this interesting issue in the future.

GRB magnetic field morphology can be investigated by polarization observation\footnote{We
note that \citet{jor19} obtained a low polarization degree of GRB 190114C in the optical band. They claim that the intrinsically low polarization
can be due to the distorted large-scale magnetic field before the reverse shock occurs.}.
If magnetic field is bipolar and large scale, linear polarization of synchrotron radiation has
high degree. If magnetic field is random, linear polarization of synchrotron radiation has
low degree \citep{gruzinov99}. We apply jitter radiation to study GRB
polarization properties, and we found that both high and low degrees can be obtained \citep{mao13,mao17}.
However, in the jitter polarization model, low luminosity GRB jets have high linear
polarization degrees, while high luminosity GRB jets have low linear polarization degrees.
This prediction is under the assumption that all GRB jets have similar view angles pointing to an
observer. We can perform the statistical analysis when we have large observational samples. The
statistical analysis can be adopted to examine the validation of the jitter radiation.

There are only a few cases of the GRB detection in the TeV energy band. We believe that MAGIC and HESS can efficiently perform GRB follow-up observations and detect more GRBs in the TeV energy band. Moreover, IACT surveys can also take a vital role on the VHE source detection. For example, large high altitude air shower observatory (LHAASO) is expected to promptly detect GRBs in the TeV energy band, and the work is ongoing. The IACT surveys can further constrain the radiation mechanisms of the VHE sources in the following years.

We suggest the jitter radiation to explain the GRB VHE emission. Moreover, we think that the jitter radiation can be also applied to other kinds of objects that have the VHE emissions. For instance, we may consider the jitter radiation to explain the blazar emission in the TeV energy band, and the radiation is dependent on the turbulent plasma properties inside the emission region. A wide application of the jitter radiation is expected in the high-energy astrophysics.

\section{Conclusions}
We propose that the GRB photons in the TeV energy band can be generated by the jitter radiation. In our scenario, kinetic turbulence takes a vital role on the production of the TeV-photons. We utilize this possibility to explain the GRB emissions recently detected by MAGIC and HESS. We expect more VHE data samples from both follow-up and survey detections by IACTs in the future.

\section*{Acknowledgements}
We appreciate the referee for the very helpful suggestions and comments. J.M. is supported by the National Natural Science Foundation of China (11673062) and the Oversea Talent Program of Yunnan Province.

\section*{Data Availability}
The data underlying this article will be shared on reasonable request to the corresponding author.










\bsp	
\label{lastpage}
\end{document}